# TASK & RESOURCE SELF-ADAPTIVE EMBEDDED REAL-TIME OPERATING SYSTEM MICROKERNEL FOR WIRELESS SENSOR NODES

Xing Kexing[1], Zuo Decheng[2], Zhou Haiying[3] and HOU Kun-Mean[4]

[1,2,3]School of Computer Science & Technology, Harbin Institute of Technology, Harbin, China
`{xingkexin,hyzhou, zdc}@ftcl.hit.edu.cn`

[4]Laboratory LIMOS CNRS 6158, University of Blaise Pascal, Clermont-Ferrand, France
`kun-mean.hou@isima.fr`

## ABSTRACT

*Wireless Sensor Networks (WSNs) are used in many application fields, such as military, healthcare, environment surveillance, etc. The WSN OS based on event-driven model doesn't support real-time and multi-task application types and the OSs based on thread-driven model consume much energy because of frequent context switch. Due to the high-dense and large-scale deployment of sensor nodes, it is very difficult to collect sensor nodes to update their software. Furthermore, the sensor nodes are vulnerable to security attacks because of the characteristics of broadcast communication and unattended application. This paper presents a task and resource self-adaptive embedded real-time microkernel, which proposes hybrid programming model and offers a two-level scheduling strategy to support real-time multi-task correspondingly. A communication scheme, which takes the "tuple" space and "IN/OUT" primitives from "LINDA", is proposed to support some collaborative and distributed tasks. In addition, this kernel implements a run-time over-the-air updating mechanism and provides a security policy to avoid the attacks and ensure the reliable operation of nodes. The performance evaluation is proposed and the experiential results show this kernel is task-oriented and resource-aware and can be used for the applications of event-driven and real-time multi-task.*

## KEYWORDS

*WSN, event-driven, thread-driven, scheduling strategy, over-the-air, security*

## 1. INTRODUCTION

Wireless sensor networks (WSN) have gained a great development in recent years. Many WSN applications are emerging rapidly and being applied in different scenarios. Unlike traditional embedded devices, the sensor nodes have many resource constraints, such as limited energy, short communication range, low bandwidth, and limited processing memory. In addition, sensor nodes are generally deployed large-scaly and far from human access. Thus, the characteristics of WSN applications bring some challenges for designing an OS on sensor nodes, described as follows:

1. To reduce power consumption and memory requirements for resource-constrained nodes.
2. To support Multi-task in view of diversity of WSN applications.
3. To update codes remotely and dynamically.
4. To provide security services in resource-constrained nodes

This paper presents a novel task and resource self-adaptive real-time microkernel for wireless sensor nodes. The kernel provides a hybrid programming model, which combines the benefits of Event-driven and Thread-driven model. Correspondingly, the kernel adopts event & thread 2 level strategies aimed at supporting real-time multi-task. Moreover, to make the nodes capable of coping with new tasks and then adjusting their behaviours in different environments, the kernel provides an over-the-air reprogramming mechanism to update the code of the OS dynamically. The kernel also provides a security service to avoid the attacks and ensure the normal operation of nodes.

The rest of this paper is organized as follows: Section 2 discusses the related work on WSN OS design. Section 3 describes the kernel in detail, including its architecture, hybrid programming model, scheduling strategy, the run-time updating mechanism and its security policy. In Section 4, we evaluate the performance and overheads, and draw the conclusions. Finally, we will give a brief conclusion in Section 5.

## 2. RELATED WORK

At present, there are two different research methodologies for WSN OS: one is to streamline the general RTOS(such as the μC/OS-II[1], VxWorks[2], Windows CE[3]) but they are not suitable for resource-constrained nodes. Another is to develop a dedicated OS according to the characteristic of WSN, such as TinyOS[4], SOS[5], MantisOS[6], and Contiki[7]. But most of dedicated WSN OS are built on network protocols without supporting multi-task and event-driven at the same time.

### 2.1. Architecture

The kernel architecture has an influence on the size of the OS kernel as well as on the way it provides services to the application programs. For resource-constrained sensor nodes, we must take the performance and flexibility into consideration at same time.

A WSN OS should have an architecture that results in a small kernel size, hence small memory footprint. The architecture must be flexible, that is, only application-required services get loaded onto the system [8].The followings are the feature analyses of three mainstream architectures [9], as described in Table 1.

Table 1. Features of 3 mainstream architectures.

| Architectures | Features |
|---|---|
| Monolithic | <ul><li>a single image for the node</li><li>not suitable for frequent changes</li></ul> |
| Modular | <ul><li>Organized as independent module</li><li>fits well in reconfiguration</li><li>extra overhead of loading and unloading modules</li></ul> |
| Virtual machine | <ul><li>whole network of nodes is a single entity</li><li>gives flexibility in developing applications</li></ul> |

## 2.2. Programming Model

An appropriate programming model not only facilitates software development but also promotes good programming practices [10]. The issue between event-driven mode and thread-driven model has been discussed in the WSN field for years. Table 2 shows the advantages and disadvantages of event-driven and thread-driven model with its corresponding OSs [11].

Table 2. Event-driven vs.Thread-driven

| Model | advantages | disadvantages | OS |
|---|---|---|---|
| Event-driven | <ul><li>Concurrency with low resources</li><li>Inexpensive scheduling</li><li>Complements the way networking protocols work</li><li>Highly portable</li></ul> | <ul><li>Event-loop is in control</li><li>Can't be preempted and do not support multi-task</li><li>Bounded buffer producer consumer problem</li></ul> | Tiny OS, SOS, Contik |
| Thread-driven | <ul><li>Eliminates bounded buffer problem</li><li>Automatic scheduling</li><li>Real-time performance</li><li>Simulates parallel execution</li></ul> | <ul><li>Complex shared memory</li><li>Expensive context switches</li><li>Complex stack analyze</li></ul> | Contiki Mantis OS |

## 2.3. Run-time Remote-updating

Due to the dense and large-scale deployment of the sensor nodes, we should provide a mechanism to update nodes over-the-air: collecting nodes to apply updates is often tedious or even dangerous [12]. Remote code-update schemes for sensor nodes meet two key challenges: (1) resource consumption and (2) integration into the system architecture. Existing approaches can be classified into three main categories.

*Full-Image Replacement*: These techniques such as XNP [14] or Deluge [15] operate by disseminating a new binary image of an application and the OS in the network. However, frequent full image replacements result in huge energy consumption and short-life of sensor node.

*Virtual Machines (VM):* VMs can reduce the energy-cost of disseminating new functionality in the network as VM code is commonly more compact than native code [16]. However, as mentioned in architecture, it is hard to be realized on a resource-constrained sensor node.

*Dynamic Module Loading*: It is also called the incremental update. Unlike the full image replacement, the kernel will not be modified and just add e new or remove modules, such as Contiki, SOS and TinyOS. This mechanism can reduce the updating code size and energy-consumption.

## 2.4. Security Guarantee

WSNs suffer from many constraints, including low computation capability, small memory, limited energy resources, susceptibility to physical capture, and the use of insecure wireless communication channels. These constraints make security in WSNs a challenge.

At present, the researches on security are classified into five categories: cryptography, key-management, secure routing, secure data aggregation, and intrusion detection. Among them, cryptography and key-management are regarded as the core issues in WSN security.

## 3. TASK & RESOURCE SELF-ADAPTIVE OPERATING SYSTEM DESIGN

### 3.1. Microkernel Design

#### 3.1.1. Architecture

As the kernel is task and resource self-adaptive, we must take the performance and the flexibility into consideration at the same time. So the kernel takes the modular structure based on its requirement as illustrated.

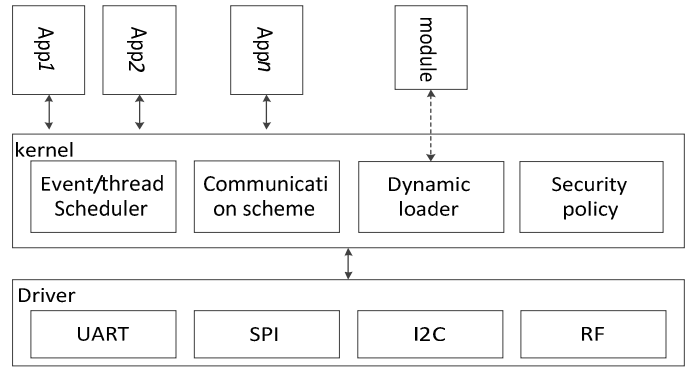

Figure 1. Architecture of task-oriented and resource-aware operating system

Figure 1 shows the kernel structure of task and resource self-adaptive WSN OS. The hardware driver is isolated from the kernel. Applications can also access kernel function by API and system call. Moreover, Applications are loaded as modules for processing the specific task. By this way, the node can adapt to the environment automatically.

#### 3.1.2. Hybrid programming model

As mentioned in section 2, the event can't be preempted and doesn't support real-time multi-task. And thread-driven model consumes much energy on context switching .The kernel integrates advantages of the event-driven and thread-driven model and proposes a hybrid programming model. The kernel is defined as a set of events, and an event is then defined as a set of threads, as shown in Equation 1 and 2

$$K=\{E_i: i=1,2,\cdots,n; \ E_1 \rightarrow E_2 \cdots \rightarrow E_n\} \qquad (1)$$

$$E_i=\{T_{i,j}: j=1,2,\cdots,n; T_{i,1}//T_{i,2}//\cdots//T_{i,n}\} \qquad (2)$$

The $K$ is an instance of the kernel, $E$ represents an event and $T$ is a thread. The symbol '→' indicates the precedence operation and '//' indicates the concurrent operation.

The feature of the hybrid programming model is that the node can easily switch operation mode between two typical programming models according to the kind of task. If we set the event number equal 1, the kernel runs in the manner of multi-thread model, shown in Figure 2**.** While if event contains a single thread, the kernel runs in the manner of event-driven model.

The kernel is task self-adaptive automatically to work in different modes, thus the scopes of applications will be greatly extended.

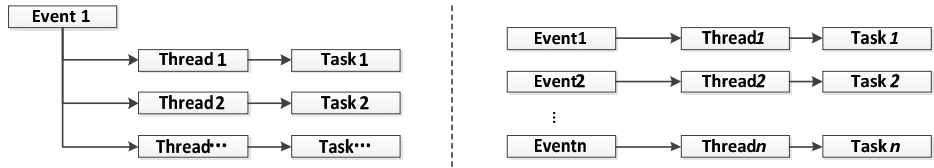

Figure 2. Switch to the thread-driven mode &Switch to Event-driven mode

### 3.1.3. Communication Scheme

From the view of operating system, the communication refers to two parts: internal communication in single node and the communication between nodes. In addition, most of WSN OS is designed for a single node without supporting communication in distributed network.

Therefore, the kernel proposes a scheme to unify the inter-process communication and communication between nodes. This scheme is based on "tuple space", a thought from a parallel programming language "Linda" [17], through "In / Out" system primitives to achieve the communication of inter-event, inter-thread, event/thread-peripherals and inter-CPU, as shown in Figure 3. Each "tuple" is identified by one numeric identifier, which can be assigned to the local or distributed buffer. And user can customize it automatically.

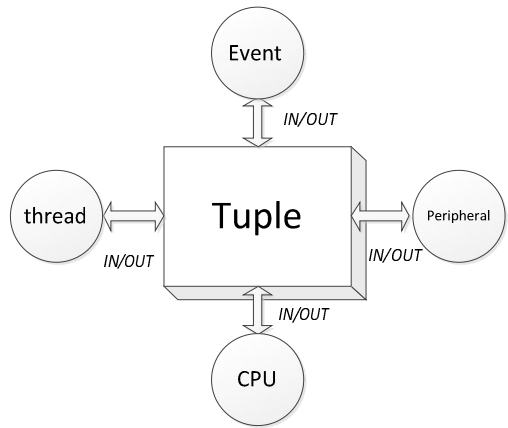

Figure 3. Tuple-based unified communication scheme

### 3.1.4. Scheduling strategy.

Based on hybrid programming model and tuple space, the kernel offers a two-level scheduling strategy: 'event-driven' (scheduling for events) and 'priority-based preemptive' (scheduling for threads).

*Event management*

Each event is associated with a unique tuple identified by a numeric identifier. A tuple has its private buffer space which is used for storing messages or signals received from hardware or software interrupts. Each event has also a priority level which is adapted automatically to the message number of the event tuple according to the event priority. The priority of the event can be pre-defined by programmer, as shown in (3)

$$Event\_next = select\_event(\ tuple\_msg\_num,\ priority) \qquad (3)$$

The strategy is sensitive to the memory consumption and proved a solution for the problem that event cant' be preempted. Thus, the kernel can make a rapid response to emergent tasks.

*Thread scheduler*

In comparison with events, threads have one more state named 'Suspend'. If the operating condition is not met, this thread is blocked and its state changes to 'Suspend'. Whereas, if the operating condition is satisfied, the suspended thread is activated to resume running and its state changes to 'Ready'. Each thread is allocated with a unique priority level. Thread can be preempted by other threads in a same event. The kernel adopts a 'priority-based preemptive' scheduling scheme and the scheduler will decide next-run thread through look up the value of *thread_priority* and *thread_timeslice*, as shown in Equation (4)

$$thread\_next = select\_thread(priotity, time\_slice) \qquad (4)$$

Through switching the threads, the system can process multiple tasks concurrently. For some periodic tasks such as data collections, sampling and interval routing, the kernel can be customized to thread-driven mode and process them at the same time. However, if the system is used to detect some events, the kernel can be switched to the event-driven mode to make a rapid response.

### 3.2. Run-time over-the-air updating mechanism

As the kernel is task-oriented, it is necessary for nodes to provide a run-time over-the-air updating mechanism, which is described from three levels according to granularity of updating code size.

*Global variable modification*

In the kernel, there exist many global variables pre-defined by developers, which can determine the operating routines. Through modifying the global variable in the ram, the node can be switched to another operation mode. So the kernel will maintain a variable table which keeps the information of the global variables, including the variable name and the physical address. After rebooting the nodes, the new mode is activated. Meanwhile, a set of remote-commands is designed to inform the node to update the global variables. The method relatively consumes little power because only several bytes are transmitted and modified.

*Dynamically loading Module*

In the modular architecture, the kernel can load new module or unload the existing module at run time without rebooting the node. Thus, just the necessary modules need to be transmitted, that is an efficient way to reduce the update code size and optimize the power usage. The procedure can be divided into the following steps [18]:

1. *Meta-Data Generation:* The meta-data consists of information about the whole program, component specific information, symbol information, and the relocation table.
2. *Code Distribution*: during this step, the new modules together with their meta-data are transmitted from the base station or the sink node.
3. *Data Storage on the Node*: Once modules with meta-data are received at the radio interface of the node, they are stored in the external FLASH memory.
4. *Register module*: the kernel merges information of the new module's meta-data with the old symbol table in order to access to the new module and invoke the new functions.
5. *Relocation Table Substitution*: the kernel will substitute the old relocation table with the new one.
6. *Update References*: through the new relocation table, the kernel can update the references of the global variable in the data segment or the function address in the code segment.

*Full image replacement*

This level is designed for updating the whole kernel image. Namely, when the new version of the kernel differs from the old one greatly, the node should erase the whole image and program the new version into the flash. For resource-constrained nodes, it is unlikely to modify the kernel frequently. So full image replacement is relatively rarely used.

## 3.3. Security policy

Due to the nature of wireless connection and unattended mode, the sensor node is prone to security attack. The kernel provides a security policy for sensor nodes, in order to guarantee the normal operation of the nodes and support reliable and efficient code distribution.

### 3.3.1. Cryptography

The traditional encryption algorithm is divided in two categories: symmetric cryptography (such as RC4, RC5, SHA-1 and MD5) and asymmetric cryptography (such as RSA and ECC). Therefore, the kernel adopts a tow-level cryptography mechanism, as shown in Figure 4. In symmetric cryptography, hashing algorithms (MD5 and SHA-11) incurred almost an order of a magnitude higher overhead than encryption algorithms (RC4, RC5). Thus, RC5 is adopted for communication between resource-constrained nodes. Meanwhile, the RSA algorithm, a typical asymmetric cryptography is applied for communication between nodes and sink node or base station which has adequate resource.

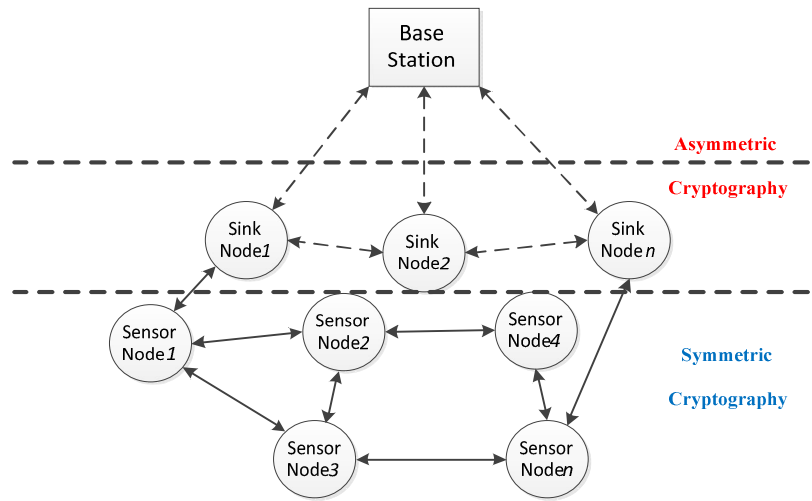

Figure 4. The two-level cryptography mechanism

### 3.3.2. Key Management.

In Sensor networks end-to-end encryption is impractical because of large number of communicating nodes and each node is incapable of storing large number of encryption keys. We assume that the number of sensor nodes in WSN is $N$. Storing the keys of other $N-1$ nodes will occupy large memory of the limited memory nodes.

In order to reduce the memory the keys occupied, the kernel takes the hop-by-hop encryption mechanism, in which each sensor node stores encryption keys shared with its immediate neighbours. Thus the keys stored in nodes can be divided into 2 parts: the keys $Kn$, which is shared with its neighbours and the key $Ks,$ *which* is shared with the base station or sink node. This mechanism reduces the memory the keys occupied and the power consumption of transmitting keys.

## 4. EVALUATION

### 4.1. Sensor platform

The kernel has been tested on the AT91SAM7S evaluation board which include AT91SAM7S256 processor core (based on arm7),256Kbytes Flash, 64Kbytes SRAM, and peripherals such as UART, SPI, and I2C etc. The RF module is XBEE-PRO [21] (based on Zigbee) which communicates with the processor core through UART (RS-232 protocol).

## 4.2. Performance analysis

### 4.2.1. Memory usage

Table 3 shows the storage usage of the kernel. It occupies no more than 5KB and is suitable for resource-constrained sensor node. The code size of whole image which includes kernel, hard drivers and applications is less than 30KB. Thus, the kernel is flexible and portable to heterogeneous sensor nodes.

Table 3. Code size and data size of the kernel and whole system

|  | Code size(bytes) | Data size(bytes) |
| --- | --- | --- |
| Kernel | 3572 | 1272 |
| Kernel+Driver+Applications | 16988 | 11929 |

### 4.2.2. Power analysis

To reduce energy consumption, the main energy-aware module of the sensor node (microprocessor, wireless RF module) support low power mode, as shown in Table 4.

Table 4. Power consumption evaluation

| Chips | | Power consumption | |
| --- | --- | --- | --- |
| | | Normal mode | Low power mode |
| AT91SAM7S256(48MHz) | | <50mA | <60μA |
| Xbee-pro | Transmit Mode | <250mA | |
| | Receive Mode | <50mA | |
| | Sleep Mode | <50μA | |

When the system is in idle state (no data transmission or no running components), the kernel can customize the node configuration to make it run at low power mode to reduce the consumption and prolong the lifetime of the node. Thus, it demonstrated that the kernel is task and resource self-adaptive for sensor node.

### 4.2.3. Overhead of scheduling strategy

The kernel adopts event/thread 2 level scheduling strategies to support real-time multi-task. This paper evaluates the overhead of scheduling strategy from 3 aspects: Interrupt response time, Event switch time and Thread switch time.

Table 5. Overhead of scheduling strategy

| | Overhead (instruction cycle) | | | Time (μS)(48Mhz) | | |
| --- | --- | --- | --- | --- | --- | --- |
| | min | avg | max | min | avg | max |
| Event switch time | 164 | 168 | 516 | 3.4 | 3.5 | 10.7 |
| Thread switch time | 176 | 93 | 798 | 3.6 | 1.9 | 16.6 |
| Interrupt response time | 157 | 194 | 395 | 3.2 | 4 | 8.2 |

Table 5 shows the overhead of the scheduling strategy through testing the average time mentioned above.

### 4.2.4. Overhead of run-time updating mechanism

As mentioned in chapter 3, the paper proposes a run-time remote-updating scheme from 3 levels according to the granularity of the updating code size. Obviously, the overhead and power consumption is determined by the code size disseminated to a great extent. The paper evaluates the overhead from two aspects, dynamic loading of modules and full image replacement.

Table 6. Comparison of energy-consumption of dynamic loading and full image replacement

| step | Dynamic loading (mJ) | Full image replacement(mJ) |
|---|---|---|
| Code size | 160bytes | 28917bytes |
| Receive data | 17mJ | 330mJ |
| Store data | 1.1mJ | 22mJ |
| Reloc&ref updating | 14.mJ | 0mJ |

As the meta-data generation is executed in sink node or base station, the paper evaluate the energy-consumption of the rest of steps, which are executed in sensor node. The table 6 shows the Comparison of energy-consumption of dynamic loading and full image replacement.

## 5. CONCLUSION

This paper presents a novel task and resource self-adaptive real-time microkernel for wireless sensor nodes. Based on the modular architecture, the kernel integrates the benefits of Event-driven and Thread-driven model to make the kernel capable of processing real-time tasks and multi-task. Moreover, in order to adapt to the complex environment, the sensor nodes are enabled to be customized and configured dynamically through the run-time over-the-air updating mechanism. A two level Cryptography mechanism and key management is also proposed for resource-constrained node to guarantee the reliable and efficient communication and code distribution.

Through analysis of the evaluation results, the kernel has good performance in communication throughout, supporting real-time task, and over-the-air updating mechanism. Compared with other dedicated WSN OSs, the kernel consumes more energy and memory. The verification of the updating code integrity should also be taken into account in over-the-air updating mechanism. In addition, there is still much room for optimizing the cryptography algorithm and managing the keys more efficiently.

Finally, the kernel extends the range of WSN application and provides the reliability and robustness for the complex application.